# Ferroelectricity in AlScN: Switching, Imprint and sub-150 nm Films

Simon Fichtner, Fabian Lofink, Bernhard Wagner
Fraunhofer ISIT
Itzehoe, Germany
simon.fichtner@isit.fraunhofer.de

Georg Schönweger, Tom-Niklas Kreutzer, Adrian Petraru, Hermann Kohlstedt
Institute for Material Science
Kiel University
Kiel, Germany

*Abstract*—The discovery of ferroelectricity in AlScN allowed the first clear observation of the effect in the wurtzite crystal structure, resulting in a material with a previously unprecedented combination of very large coercive fields (2-5 MV/cm) and remnant polarizations (70-110 µC/cm²). We obtained initial insight into the switching dynamics of AlScN, which suggests a domain wall motion limited process progressing from the electrode interfaces. Further, imprint was generally observed in AlScN films and can tentatively be traced to the alignment of charged defects with the internal and external polarization and field, respectively. Potentially crucial from the application point of view, ferroelectricity could be observed in films with thicknesses below 30 nm – as the coercive fields of AlScN were found to be largely independent of thickness between 600 nm and 27 nm.

I. INTRODUCTION

The drive towards miniaturization of piezoelectric sensors and actuators as well as the introduction of ferroelectric functionality into integrated circuit technology has led to substantial scientific and commercial interest in ferroelectric thin films. Many of the more important ferroelectrics are perovskite oxides, with typical disadvantages such as low transition temperatures, limited reliability and stability, non-linear displacement or compatibility issues with complementary metal-oxide-semiconductor technology (CMOS). Therefore, novel ferroelectric materials with improved properties could lead to substantial technological innovation.

Here we give an overview of the recent discovery of ferroelectricity in CMOS compatible, III-N semiconductor based AlScN [1]. Next to highlighting the profound extension that AlScN provides to the ferroelectric parameter range available to thin film technology, this paper provides first insight into the switching dynamics of the material, imprint and ferroelectric films with thicknesses below 150 nm.

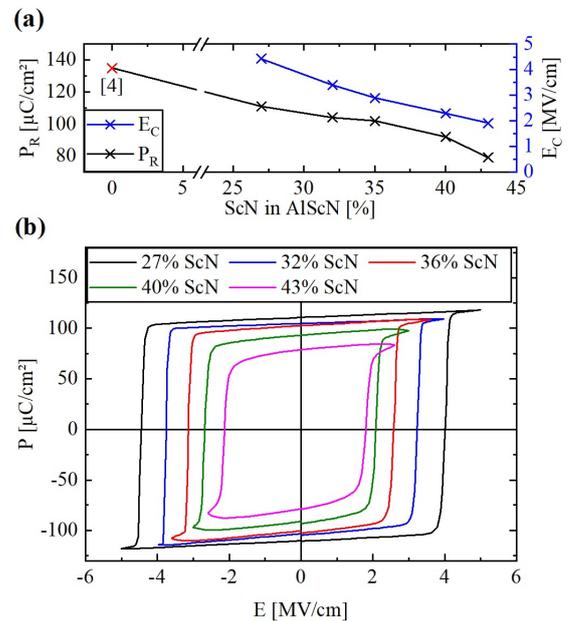

Figure 1: (a) Remnant polarization $P_R$ and coercive field $E_C$ over ScN content in AlScN. (b) Corresponding P-E hysteresis loops measured at 711 Hz.

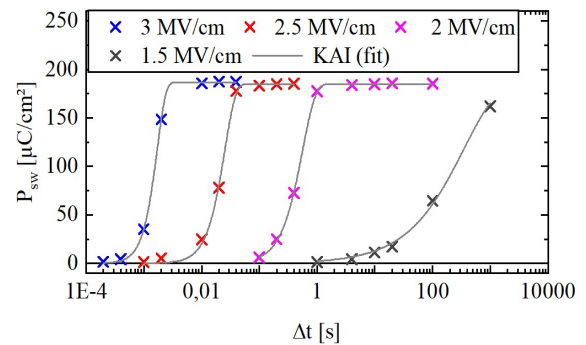

Figure 2: Switched polarization $P_{sw}$ over pulse width (triangular) for different electric field amplitudes with fitted curves in accordance with the KAI model [5]. Measured on $Al_{0.64}Sc_{0.36}N$.

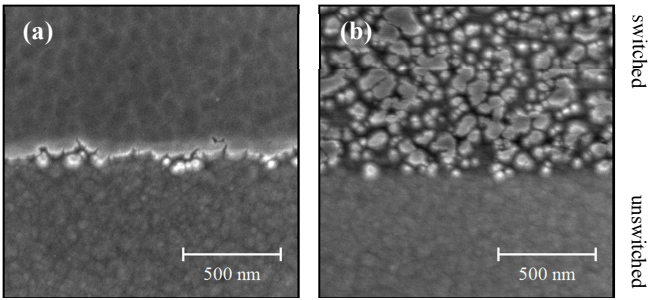

Figure 3: SEM top view of the step between switched and unswitched $Al_{0.64}Sc_{0.36}N$ after removal of the top electrode and brief etching in $H_3PO_4$. (a) Switched with an electric field amplitude $E_{max}/E_C = 1$ and (b) switched with $E_{max}/E_C = 0.93$.

## II. METHODS/RESULTS

AlScN thin films with thicknesses between 27 nm and 1 μm were deposited by single and dual target sputtering as described before [1], [2]. For the top and bottom electrode, 100 nm of Pt was used, the former being structured by ion beam etch to define the capacitor structures. Characterization of the ferroelectric properties was performed with an AixACCT TF 2000 analyzer. Estimates for the AlScN film thickness were obtained from SEM images of the film cross sections, while the material composition of a particular process setting was determined through SEM EDX of films with 500-1000 nm thickness.

Exemplary P-E loops of AlScN are given in Figure 1 for Sc contents between 27% and 43% ScN in AlScN – next to their mean coercive fields $E_C$ and remnant polarizations $P_R$ over the Sc content.

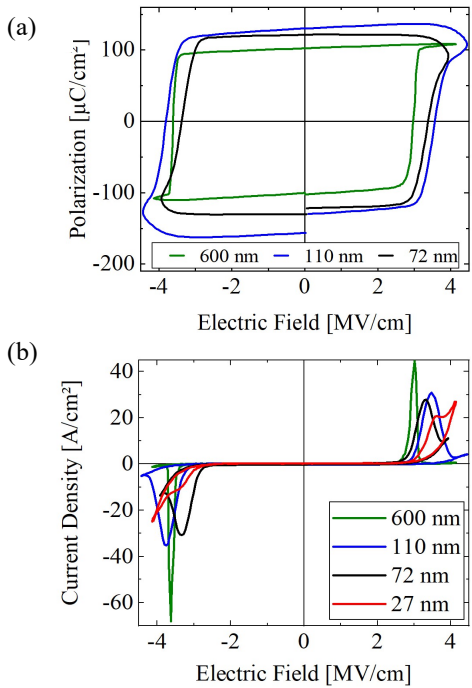

Figure 4: (a) Measured P-E loops of $Al_{0.65}Sc_{35}N$ films with thicknesses between 600 nm and 72 nm. (b) Underlying displacement currents of the measurements in part (a) plus displacement current of an $Al_{0.65}Sc_{0.35}N$ film with 27 nm thickness.

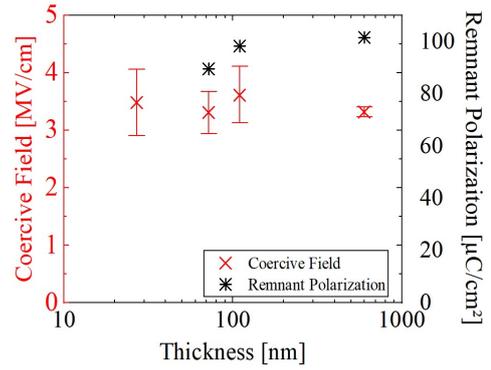

Figure 5: Coercive field and remnant polarization of $Al_{0.65}Sc_{0.35}N$ over film thickness.

Figure 2 shows pulse switching measurements with varied electric field strength and pulse length. In order to visualize the underlying domain states, brief (5 sec) anisotropic etching of the film surface was performed on capacitor structures previously exposed to switching at varied ratios of $E_{max}/E_C$, where $E_{max}$ is the amplitude of the applied electric field. For $E_{max}/E_C = 1$ and 0.93, SEM images of the resulting surface are given in Figure 3.

P-E hysteresis loops of films with thicknesses below 150 nm are given in Figure 4 – next to their displacement current. Figure 5 gives a comparison of the coercive field and remnant polarization of the same films. For the latter, the coercive field was extracted from the displacement current in order to be able to make a reasonable estimate also for the 27 nm thick film. Error bars are in accordance with the standard deviation of the individual thickness measurements. In addition, a reduced measurement frequency of 2 kHz was chosen for the 27 nm sample in order to get a clearer separation between leakage and switching currents – compared to 5 kHz for the thicker ones.

The remnant polarization was calculated by taking the slope of the linear, low field regime of the P-E loop with an axis intersection that results in this slope passing the hysteresis loop at its maximum field. Unlike for films with thicknesses of at least several 100 nm, attempts of stress control for thinner films did not result in the expected results – with all sub-150 nm films investigated so far being under compressive stress of at least 500 MPa. This was determined by differential profilometer based wafer-curvature measurements (only films with thicknesses of above 50 nm were investigated).

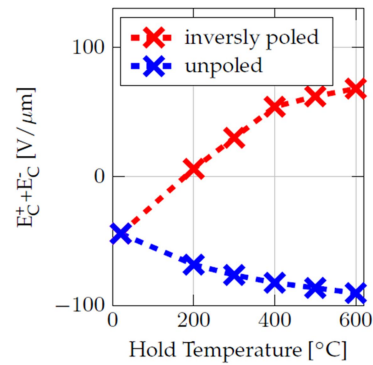

Figure 6: Center field of the P-E loop obtained from a single $Al_{0.64}Sc_{0.36}N$ capacitor after cumulative exposures of 5 minutes at the stated temperatures.

Figure *6* compares the evolution of imprint with added thermal energy between an N-polar, as deposited capacitor and a metal-polar capacitor. Imprint is expressed as the centre field of the P-E loop. The cumulative ex-situ temperature treatments were performed at ambient atmosphere

III. DISCUSSION/INTERPRETATION

While the binary wurtzite semiconductors are considered to be purely pyroelectric, increasing the foreign component inside wurtzite solid solutions as well as tensile stress applied to the wurtzite basal plane can lower the energy barrier between the polarization states of the structure sufficiently that clear ferroelectricity can be observed in AlScN (Fig. 1) [1], [3]. Contrary to ferroelectricity in e.g. hafnia based compounds, ferroelectricity in the wurtzite structure is therefore not related to the energetic stabilization of an additional polar phase but rather to the destabilization of the native phase. Besides allowing the experimental observation of ferroelectricity in AlScN, adjusting the Sc content as well as the in plane stress allows systematic control of the spontaneous (remnant) polarization $P_r$ and $E_C$ in broad intervals (Fig. 1). With its particularly large spontaneous polarization and coercive fields, AlScN is able to greatly expand the range of ferroelectric parameters that is available to microelectronics and microelectromechanical systems – as illustrated in the comparison with established materials given in Figure 7. CMOS compatibility, good temperature stability ($T_C > 600°C$), moderate deposition temperatures (300 - 400°C), excellent retention and broad linear displacement regimes add to the attractiveness of AlScN [1].

As the first wurtzite material with clear ferroelectric properties, AlScN was further able to solve the uncertainty related to the polarization constant of this particular crystal class which arises from the implementation of different reference structures in the context of the modern theory of polarization [4]. It was confirmed that the wurtzite polarization constant can reach values larger than 100 µC/cm².

Pulse switching measurements indicate a domain wall motion limited switching process, resulting in well delimited curves for varied field strength – potentially making AlScN particularly suitable for crossbar related architectures. Narrower switching intervals are obtained for higher electric fields. Fitting the KAI model to the switching curves in Figure *2* correspondingly suggests an increase of the dimension of domain propagation with increasing field strength [5].

Anisotropic etching reveals a closed, single domain surface near the top electrode down to $E_{max}/E_C = 1$. This implies that switching originates near the electrode interface(s) (Figure *3*) – since less material has to be switched in the bulk or near the bottom electrode in a case were the overall polarization was not driven into saturation. For lower ratios of $E_{max}/E_C$, stable regions with inverse polarization are also found close to the top electrode.

AlScN films were observed to generally feature polarization imprint in their as deposited state. No strong dependency on particular electrode materials or the symmetry of the compensator could be identified so far. The imprint in as deposited films always favours the N-polar state – i.e. results in the negative coercive field having a larger absolute value than the positive one. While imprint is found to become reduced under repeated cycling of the material, temperature and the majority polarization of the film exert a profound effect on its sign and magnitude. The measurement summarized in Figure *6* suggests that imprint appears due to alignment of charged defects with the majority polarization – similar to what is attributed to oxygen vacancies in PZT [6]. Therefore, changing the majority polarization from N-polar to metal polar tends to slowly realign these defects – which can be greatly sped up by providing thermal energy. This effect even allows to almost completely reverse the level of imprint, shifting the hysteresis centre from negative electric fields to positive fields. In concert with the internal electric field, application of strong external electric fields can also affect the magnitude of imprint – explaining why, under repeated cycling, imprint can virtually vanish. The underlying defect species has yet to be determined, however.

Perhaps the most crucial aspect when considering the potential of AlScN for integrated circuits such as memories is the scaling of its ferroelectric properties with film thickness. Our initial characterization series of $Al_{0.65}Sc_{0.35}N$ films allows the conclusion that ferroelectricity can be found in films with less than 30 nm thickness. Unlike typically the case for e.g. perovskite ferroelectrics, the coercive field of AlScN has the favourable property to remain rather constant over the whole investigated thickness range (Figure 5). While a sizeable polarization can be observed down to 27 nm, leakage precludes a quantitative statement on the evolution of the polarization. Different than the breakdown field, which was comparable between films of different thickness, the leakage currents observed before breakdown increased towards thinner films (Figure 4 (b)). Therefore, methods to reduce the charge leakage by engineering e.g. the layers in contact with the AlScN interfaces should be studied in the future. Besides this, films with higher Sc contents and, potentially more feasible, tensile films should be able to increase the gap between coercive field and excessive leakage as well as breakdown further and will be the focus of future studies.

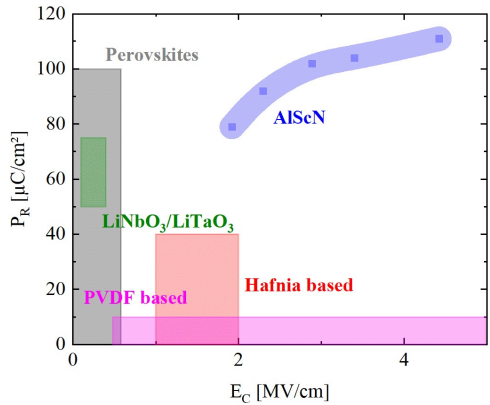

Figure 7: Remnant polarization $P_R$ over coercive field $E_C$ for common ferroelectric material classes compared to AlScN

IV. CONCLUSIONS

AlScN can provide a profound addition to the ferroelectric parameter intervals available to microtechnology – in concert with good compatibility and stability. Therefore, AlScN has the potential to greatly expand both the availability of ferroelectric films in microtechnology and the number of applications that could be potentially addressed. Regarding the switching behaviour of AlScN, our initial measurements suggest that the polarization inversion progresses from the (top) electrode in domain wall motion limited fashion. While imprint can be generally observed in the material, its sign and magnitude can be influenced by the majority polarization, external fields, and temperature – which can be tentatively explained by the presence and alignment of charged defects.

The observation of ferroelectric AlScN films with thicknesses of significantly below 100 nm was largely made possible by a favourable scaling behaviour of the coercive field in the material – a feature that might prove crucial for the eventual realization of AlScN based ferroelectric microelectronic applications. To further the application driven development of ferroelectric AlScN, future studies should focus on reducing both leakage and the coercive field, e.g. through strain and interface engineering.

ACKNOWLEDGMENT

The authors acknowledge funding by the German Federal Ministry of Education and Research (BMBF) under grant 16ES1053.